\documentclass[amsmath,amssymb,pra,final,showpacs,twocolumn]{revtex4-2}

\usepackage{amssymb,amsmath} 
\usepackage{graphicx,epsfig} 

\newcommand {\mbf}[1]{{\mathbf{#1}}}

\newcommand{\cm}{\mathrm{c\!\:\!.m\!\:\!.}}

\begin{document}

\title{Nonlocal interaction and collective excitation in deuteron breakup
  on ${}^{24}$Mg nucleus}


\author{A. Deltuva}
\email{arnoldas.deltuva@tfai.vu.lt}
\affiliation
{Institute of Theoretical Physics and Astronomy, 
Vilnius University, Saul\.etekio al. 3, LT-10257 Vilnius, Lithuania
}

\author{D. Jur\v{c}iukonis}
\affiliation
{Institute of Theoretical Physics and Astronomy, 
Vilnius University, Saul\.etekio al. 3, LT-10257 Vilnius, Lithuania
}

\received{June 5, 2024}

\begin{abstract}%
Deuteron breakup in collision with a ${}^{24}\mathrm{Mg}${} nucleus
is studied using rigorous three-body scattering equations,
extended to include also the  excitation of the nucleus. Predictions based on
local and nonlocal nucleon-nucleus optical   potentials
  with rotational quadrupole deformation enabling the excitation 
of the ${}^{24}\mathrm{Mg}(2^+)$ state are compared. 
The nonlocality effect is less pronounced
than in the deuteron inelastic scattering
${}^{24}\mathrm{Mg}(d,d')${} at the same energies, and manifests itself 
quite differently for semi-inclusive differential cross sections
of elastic and inelastic breakup.
\end{abstract}

\maketitle

\section{ Introduction} 

The description of few-cluster direct nuclear reactions has a long history.
It started with the optical potentials of a simple local Woods-Saxon form  to model the 
nucleon-nucleus interactions
and  with various distorted-wave  approximations to describe the scattering states.
Over the years  more sophisticated methods to solve the quantum-mechanical few-body problem
have been developed, such as the 
continuum-discretized coupled channels (CDCC) method  \cite{austern:87}
and the rigorous Faddeev scattering theory \cite{faddeev:60a,alt:67a}.
The interactions of a more elaborated form have been proposed as well, such as nonlocal 
\cite{pereybuck,giannini} and microscopically motivated ones \cite{furumoto:mgop,microOP:23},  
and taking explicitly into account the internal degrees of freedom of the involved nuclei,
i.e., their collective excitations, most often simulated by the rotational or vibrational models
of the core excitation \cite{tamura:cex,bohr-motelson,thompson:88}.
Remarkably, only recently those achievements in Faddeev-type calculations and nonlocal interaction
models with the core excitation have been combined together.
The studied  examples were the deuteron stripping and pickup reactions
${}^{10}\mathrm{Be}(d,p){}^{11}\mathrm{Be}$ and ${}^{11}\mathrm{Be}(p,d){}^{10}\mathrm{Be}$
\cite{deltuva:23a} and the inelastic deuteron scattering   ${}^{24}$Mg$(d,d')$ 
\cite{deltuva:23b}. 
In both beryllium and magnesium cases  the consistency
between the two-body  and three-body description
and the achieved agreement with the experimental data 
was considerably improved as compared to previous studies, presumably 
due to more sophisticated potentials and treatment of the three-body dynamics.
Furthermore, the study \cite{deltuva:23b} explicitly compared predictions with local
and nonlocal potentials and revealed significant nonlocality effects in the
inelastic deuteron scattering ${}^{24}\mathrm{Mg}(d,d')$ leading to the first excited $2^+$ state of  ${}^{24}$Mg.
Thus, it would be interesting to evaluate
the importance of the nonlocality in other channels of the $d+{}^{24}\mathrm{Mg}$ collision
such as the deuteron breakup channels.
In particular, the deuteron breakup reaction with the simultaneous excitation of the nuclear
core has not yet been studied using the Faddeev-type formalism. Therefore in the present work
we aim to investigate the $d+{}^{24}\mathrm{Mg}$ scattering leading to the breakup of the deuteron,
both with and without the excitation of the ${}^{24}$Mg nucleus.
We take over from Ref.~\cite{deltuva:23b}
the local and nonlocal nucleon-nucleus interaction models with the core excitation
and extent the three-body reaction calculations of  Ref.~\cite{deltuva:23b} to the
breakup amplitudes.

\section{Scattering equations and breakup amplitudes} 

We describe the deuteron-nucleus scattering using the
Alt, Grassberger, and Sandhas (AGS) equations  \cite{alt:67a}
 for three-particle transition operators
\begin{equation}  \label{eq:Uba}
U_{\beta \alpha}  = (1-{\delta}_{\beta\alpha}) \, G^{-1}_{0}  +
\sum_{\gamma=1}^3   (1-{\delta}_{\beta \gamma}) \, t_{\gamma} 
\, G_{0} U_{\gamma \alpha}.
\end{equation}
Here Greek subscripts label the spectator particle (or, equivalently, the pair in 
the odd-man-out notation), 
$G_0$ is the free resolvent,  and  $t_{\gamma}$
is the two-particle transition operator for the pair $\gamma$.
In the following the nucleus, proton, and neutron will be labeled by $A$, $p$, and $n$,
respectively.

When the excitation of the nucleus $A$ is included, the operators in Eq.~(\ref{eq:Uba})
become multi-component operators whose components in our notation will be distinguished
by  additional superscripts  corresponding to the ground or excited internal state of $A$,
taking value 0 or 1, respectively. The free resolvent is diagonal in the core state  label, i.e.,
\begin{subequations}  \label{eq:g0}   
\begin{align}  
 G_0^0 = {}& (E+i0  - K)^{-1}, \\
 G_0^1 = {}& (E+i0 -  \Delta m_A - K)^{-1},
\end{align}
\end{subequations}
with $E$ being the available energy in the center-of-mass (c.m.) frame, $K$ being the kinetic energy operator,
and $\Delta m_A$ being the core excitation energy,
equal to 1.369 MeV for the considered first $2^+$ excited state of  the ${}^{24}$Mg nucleus.

The neutron-proton potential $v_A$ and transition operator $t_A$ obviously cannot excite the core, having
the uncoupled components
\begin{subequations}  \label{eq:ta}   
\begin{align}  
 t_A^0 = {}&  v_A +  v_A G_0^0 t_A^{0}, \\
 t_A^1 = {}&  v_A +  v_A G_0^1 t_A^{1}.
\end{align}
\end{subequations}

The core excitation proceeds via its interaction with any of the nucleons. Therefore
the nucleon-nucleus transition operators $t_n$ and $t_p$ couple the ground and excited states 
of the core nucleus as the respective potentials $v_n$ and $v_p$ do, i.e.,
\begin{subequations}  \label{eq:tn}   
  \begin{align}
t_n^{00} = {}& v_n^{00} +  v_n^{00} G_0^0 t_n^{00} +  v_n^{01} G_0^1 t_n^{10}, \\
t_n^{10} = {}& v_n^{10} +  v_n^{10} G_0^0 t_n^{00} +  v_n^{11} G_0^1 t_n^{10}, \\
t_n^{01} = {}& v_n^{01} +  v_n^{00} G_0^0 t_n^{01} +  v_n^{01} G_0^1 t_n^{11}, \\
t_n^{11} = {}& v_n^{11} +  v_n^{10} G_0^0 t_n^{01} +  v_n^{11} G_0^1 t_n^{11}.
\end{align}
\end{subequations}
The corresponding system for $t_p$ is obtained simply replacing the subscript $n$  by $p$ in 
Eqs.~(\ref{eq:tn}).

We consider the breakup of deuteron in the collision with the ${}^{24}$Mg nucleus. Thus, the initial state
is the deuteron plus nucleus state $|\phi_d \, \mbf{q}_A^0\rangle$, a product of the 
 deuteron bound-state wave function $\phi_d$ and   the plane wave
for the relative nucleus-deuteron motion with momentum $\mbf{q}_A^0$;
the dependence on the discrete quantum numbers is suppressed in our notation.
The relevant set of the  three-body  transition operators in the AGS equations
obeys the coupled system
\begin{subequations}  \label{eq:AGScpl}   
\begin{align}  
U_{AA}^{00} = {}& t_p^{00} G_0^0 U_{pA}^{00} + t_n^{00} G_0^0 U_{nA}^{00} + 
t_p^{01} G_0^1 U_{pA}^{10}  +      t_n^{01} G_0^1 U_{nA}^{10}, \\
U_{AA}^{10} = {}& t_p^{10} G_0^0 U_{pA}^{00} + t_n^{10} G_0^0 U_{nA}^{00} + 
t_p^{11} G_0^1 U_{pA}^{10}  +      t_n^{11} G_0^1 U_{nA}^{10}, \\
U_{pA}^{00} = {}& ({G_0^0})^{-1} + t_n^{00} G_0^0 U_{nA}^{00} + 
t_A^{0} G_0^0 U_{AA}^{00} +  t_n^{01} G_0^1 U_{nA}^{10}, \\
U_{pA}^{10} = {}& t_n^{10} G_0^0 U_{nA}^{00} + 
t_A^{1} G_0^1 U_{AA}^{10} +  t_n^{11} G_0^1 U_{nA}^{10}, \\
U_{nA}^{00} = {}& ({G_0^0})^{-1}  + t_A^{0} G_0^0 U_{AA}^{00} + 
t_p^{00} G_0^0 U_{pA}^{00} +  t_p^{01} G_0^1 U_{pA}^{10}, \\
U_{nA}^{10} = {}&  t_A^{1} G_0^1 U_{AA}^{10} + 
t_p^{10} G_0^0 U_{pA}^{00} +  t_p^{11} G_0^1 U_{pA}^{10}.
\end{align}
\end{subequations}
Equations  (\ref{eq:AGScpl}) are solved in the momentum-space partial wave representation.
The on-shell matrix elements of transition operators (\ref{eq:AGScpl}) yield amplitudes
for elastic, inelastic, and rearrangement collisions\cite{deltuva:23a,deltuva:23b}.
The breakup operators are those with $\beta=0$ in Eq.~(\ref{eq:Uba}) and
are obtained from  operators (\ref{eq:AGScpl}) as quadratures
\begin{subequations}  \label{eq:AGS0}   
\begin{align}  
  U_{0A}^{00} = {}& (G_0^0)^{-1} + t_A^{0} G_0^0 U_{AA}^{00} + t_p^{00} G_0^0 U_{pA}^{00} + t_n^{00} G_0^0 U_{nA}^{00}
  \nonumber  \\ {}&
  + t_p^{01} G_0^1 U_{pA}^{10}  +      t_n^{01} G_0^1 U_{nA}^{10}, \\
U_{0A}^{10} = {}& t_p^{10} G_0^0 U_{pA}^{00} + t_n^{10} G_0^0 U_{nA}^{00}
\nonumber  \\ {}&
+  t_A^{1} G_0^1 U_{AA}^{10} + 
t_p^{11} G_0^1 U_{pA}^{10}  +      t_n^{11} G_0^1 U_{nA}^{10}.
\end{align}
\end{subequations}
The final breakup state $|\mbf{p}\mbf{q}\rangle _\alpha^c$
can be characterized by the internal state $c$ of the core 
and two Jacobi momenta for the relative motion of three free particles; any of the three Jacobi configurations 
$\alpha$ can be used equally well.
Therefore the amplitude for the deuteron breakup  leading
to the state $c$ of the final nucleus is given by
\begin{equation} 
  \mathcal{T}^c_A(\mbf{p}_\alpha,\mbf{q}_\alpha) =
 {}_\alpha^c \! \langle \mbf{p}\mbf{q} | U_{0A}^{c0} |\phi_d \, \mbf{q}_A^0\rangle.
\end{equation}

The exclusive or semi-inclusive  differential  cross section for the 
three-cluster breakup is obtained by partial integration of the standard relation 
\begin{gather}  \label{eq:d6s}
\begin{split}
d^6\sigma^c = {} & {} (2\pi)^4 \, \frac{M_A}{q_A^0} \,
\delta \left(E-\frac{p_\alpha^2}{2\mu_\alpha} - \frac{q_\alpha^2}{2M_\alpha} \right) \,
\\ & \times
|\mathcal{T}^c_A(\mbf{p}_\alpha,\mbf{q}_\alpha)|^2 \,
d^3\mbf{p}_\alpha d^3\mbf{q}_\alpha,
\end{split}
\end{gather}
where $\mu_\alpha$ and $M_\alpha$ are the reduced masses for the pair and spectator
$\alpha$, respectively. In the present work we will show results for the semi-inclusive
angular cross section in the c.m. frame
\begin{gather}  \label{eq:d2s}
  \frac{d\sigma^c}{d\Omega_\alpha}
= (2\pi)^4 \, \frac{M_A}{q_A^0} \, \mu_\alpha  \, 
\int d^2\mbf{\hat{p}}_\alpha \, d q_\alpha \, {p_\alpha} q_\alpha^2
|\mathcal{T}^c_A(\mbf{p}_\alpha,\mbf{q}_\alpha)|^2.
\end{gather}
Here the magnitude of the relative pair momentum
${p_\alpha} = \sqrt{2\mu_\alpha(E-q_\alpha^2/2M_\alpha)}$
is determined by the energy conservation that is reflected by the
$\delta$-function in Eq.~(\ref{eq:d6s}).

Another characteristic observable is the energy distribution of the selected particle
in the c.m. frame, obtainable from  Eq.~(\ref{eq:d6s}) as
\begin{gather}  \label{eq:d2se}
  \frac{d\sigma^c}{dE_\alpha}
= (2\pi)^4 \, \frac{M_A}{q_A^0} \, \mu_\alpha  \, m_\alpha \,
\int d^2\mbf{\hat{p}}_\alpha \, d^2\mbf{\hat{q}}_\alpha \, {p_\alpha} q_\alpha
|\mathcal{T}^c_A(\mbf{p}_\alpha,\mbf{q}_\alpha)|^2,
\end{gather}
with $q_\alpha = \sqrt{2m_\alpha \,E_\alpha}$ and particle mass $m_\alpha$.

The AGS equations are formally applicable only to systems with short-range 
interactions as assumed above.  Nevertheless, the inclusion of the long-range Coulomb force 
is possible using the method of screening and renormalization
\cite{taylor:74a,semon:75a,alt:80a,deltuva:05a}.
The nuclear potential is supplemented by  a screened Coulomb one,  
the AGS equations are solved, and 
the resulting amplitudes are  renormalized in the unscreened limit, leading to proper
physical amplitudes. The essential point of this procedure is that screening radius of the
order of 10 fm is sufficient for the
 short-range part of the amplitude to reach the unscreened limit with good accuracy.


 \section{Nonlocal potential with nuclear excitation}

 In the momentum-space solution of the three-body scattering problem the inclusion of nonlocal potentials
 does not lead to additional technical complications  as compared to local ones.
 We take the proton-${}^{24}$Mg and neutron-${}^{24}$Mg optical potentials from
 Ref.~\cite{deltuva:23b}, i.e.,  
 \begin{equation}  \label{eq:Vdj}
   v_N(\mbf{r}_f,\mbf{r}_i;\hat{\xi}) =
   \frac12 \big[ V(r_f,\hat{\xi}) H(r_{-}) + H(r_{-}) V(r_i,\hat{\xi}) \big].
\end{equation}
 Here $\mbf{r}_i$ and $\mbf{r}_f$ are the initial and final distances between particles,
 $\hat{\xi}$ refers to the internal nuclear degrees of  freedom in the body-fixed frame
 \cite{tamura:cex,thompson:88}, $\mbf{r}_{-} = \mbf{r}_f - \mbf{r}_i$, 
  $V(r,\hat{\xi})$ is a local potential with core excitation, and
 \begin{equation}  \label{eq:Hx}
 H(r_{-}) = \pi^{-3/2} \rho^{-3} e^{-(r_-/\rho)^2}
\end{equation}
 is the nonlocality function with the nonlocality range $\rho$.
 
 The local potential $ V(r,\hat{\xi})$ is a standard optical potential with
 real volume and spin-orbit as well with imaginary volume and surface terms,
 extended to include the quadrupole rotational core deformation \cite{tamura:cex,thompson:88}.
 Taking the nonlocality $\rho = 1$ fm, 
 the other parameters were determined in Ref.~\cite{deltuva:23b}
 by fitting the experimental $p+{}^{24}$Mg data \cite{pmgc} for the elastic and inelastic ($2^+$)
 differential cross section at proton beam energies $E_p = 30.4$, 34.9, 39.9 and 44.9 MeV.
Due to absence of accurate neutron scattering data the same parameters (without Coulomb) were taken 
for the neutron-nucleus potential as justified in Ref.~\cite{deltuva:23b}.
 In addition, nearly-equivalent local potentials were developed in order to evaluate the nonlocality effect. 
 The nonlocal potential was able to fit the two-body data in a broader energy range
 and provided a better description of the three-body data \cite{mg:d}, especially for the
 deuteron-nucleus inelastic scattering at intermediate angles \cite{deltuva:23b}.
 In the present work we will use local and nonlocal potentials from  Ref.~\cite{deltuva:23b}
 to study the nonlocality effect in the deuteron breakup reaction. Since 
five sets of nonlocal potential parameters were given in Table I of  Ref.~\cite{deltuva:23b}, their predictions
 will be collected into bands as done in Ref.~\cite{deltuva:23b}.
As one more measure for the uncertainty related to different parameter sets we calculated the proton-nucleus reaction
cross section that was not constrained by the fit. It has a typical $\pm 3\%$ spread, e.g.,   $\sigma_R = 635 \pm 20$ mb at 34.9 MeV.
 The sensitivity of the results to the choice of a realistic  neutron-proton interaction is low;
the presented results are obtained using the CD Bonn potential \cite{machleidt:01a}.

\section{Results for the semi-inclusive deuteron breakup}

In the previous work \cite{deltuva:23b} we studied
differential cross sections for the elastic and inelastic
$d+{}^{24}$Mg scattering  at deuteron beam energies  $E_d = 60$, 70, 80 and 90 MeV.
The most spectacular nonlocality effect was observed at $E_d = 90$ MeV. On the other hand,
in  the nucleon-nucleus subsystem the best agreement between local and nonlocal optical potentials
was achieved near 35 MeV, half of $E_d = 70$ MeV. In order to cover these extreme cases here
we present the differential cross section for the deuteron breakup on the ${}^{24}$Mg nucleus
at  $E_d = 70$ and 90 MeV. As it is quite common in the literature, we refer to
deuteron breakup reactions with the final ${}^{24}$Mg nucleus being in
its ground (excited) state as the elastic (inelastic) breakup.

\begin{figure}[!]
\includegraphics[scale=0.90]{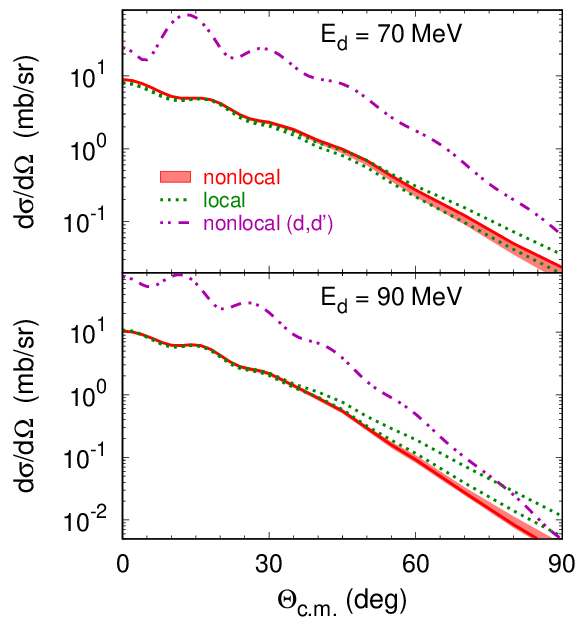}
\caption{\label{fig:i} (Color online) Semi-inclusive 
differential cross section for the inelastic deuteron breakup on the ${}^{24}$Mg nucleus
at  $E_d = 70$ and 90 MeV as a function of the ${}^{24}$Mg($2^+$) scattering angle $\Theta_{\cm}$
in the  c.m. frame.
Predictions obtained with different parameter sets of the nonlocal optical potential
from  Ref.~\cite{deltuva:23b}
  are combined into bands, while dotted curves represent predictions
  based on  local potentials from  Ref.~\cite{deltuva:23b}.
  The double-dotted dashed curves show the differential cross section
  for the ${}^{24}$Mg$(d,d')$ reaction with a single parameter set of the nonlocal optical potential
  taken from Ref.~\cite{deltuva:23b}.
 }
\end{figure}

In Fig.~\ref{fig:i} we present the semi-inclusive
differential cross section (\ref{eq:d2s}) for the inelastic breakup.
The observable is shown as a function of the nucleus
scattering angle  $\Theta_{\cm}$ in the  c.m. frame, which is the same kinematic variable
as used for the description of elastic and inelastic deuteron-nucleus scattering in
Figs.~3 and 4 of Ref.~\cite{deltuva:23b}, except that here the energies of the final-state particles
are not fixed but integrated over according to Eq.~(\ref{eq:d2s}). Nevertheless, both
differential cross sections of this work and of  Ref.~\cite{deltuva:23b} characterize the probability
for a given nucleus scattering angle in the c.m. frame. It is therefore interesting to compare them
as done in Fig.~\ref{fig:i}. Remarkably, despite  that the available energy is well above the breakup threshold,
the deuteron breakup with a simultaneous core excitation is significantly less probable
than the inelastic scattering of the deuteron. The $(d,d')$ cross section shows
also a more rich angular structure.
The nonlocality effect for the deuteron inelastic breakup appears
less significant than the one observed
in the inelastic deuteron scattering and shown in Fig.~3  of Ref.~\cite{deltuva:23b}.
In Fig.~\ref{fig:i} the two sets of predictions partially overlap almost in the whole
angular regime except for large angles at $E_d = 90$ MeV.

\begin{figure}[!]
\begin{center}
\includegraphics[scale=0.90]{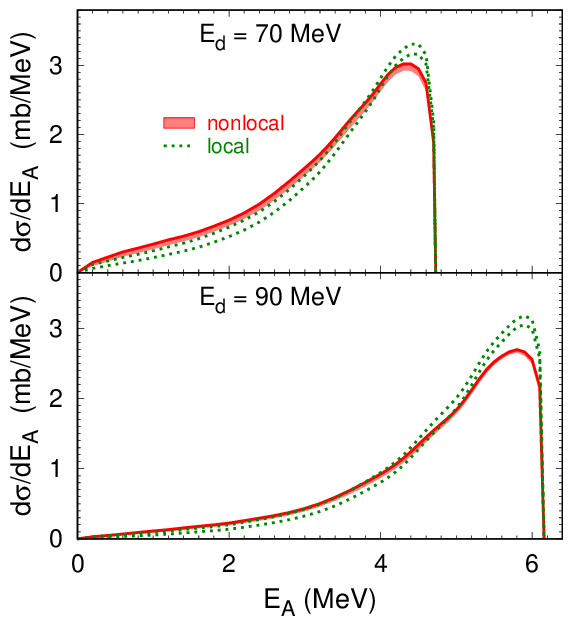}
\end{center}
\caption{\label{fig:ie} (Color online) Semi-inclusive
differential cross section for the deuteron inelastic breakup on the ${}^{24}$Mg nucleus
at  $E_d = 70$ and 90 MeV as a function of the ${}^{24}$Mg$(2^+)$ energy $E_A$
in the  c.m. frame.
Predictions obtained with different parameter sets of the nonlocal optical potential
from  Ref.~\cite{deltuva:23b}
  are combined into bands, while dotted curves represent predictions
  based on  local potentials from  Ref.~\cite{deltuva:23b}.
}
\end{figure}

The energy distribution (\ref{eq:d2se}) of the ${}^{24}$Mg$(2^+)$ nucleus presented in  Fig.~\ref{fig:ie}
reveals more significant systematic differences. The nonlocality decreases the peak near the maximal
allowed energy and increases the differential cross section in the lower-energy tail. Due to 
the opposite sign of the effect in those two regions the integrated cross section for the
inelastic breakup (see also Table \ref{tab:s}) is almost unaffected by the nonlocality. 

\begin{figure}[t]
\begin{center}
\includegraphics[scale=0.90]{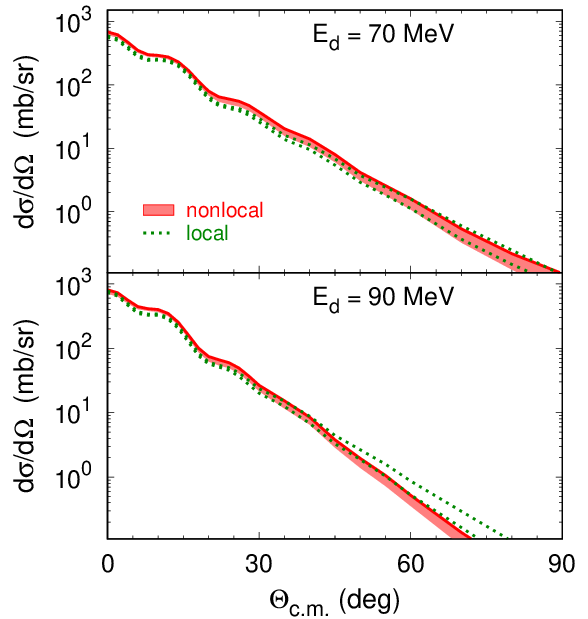}
\end{center}
\caption{\label{fig:e} (Color online) Semi-inclusive
differential cross section for the deuteron elastic breakup on the ${}^{24}$Mg nucleus
at  $E_d = 70$ and 90 MeV as a function of the ${}^{24}$Mg scattering angle $\Theta_{\cm}$
in the  c.m. frame. 
Curves and bands are as in Fig.~\ref{fig:ie}.
}
\end{figure}

\begin{figure}[!]
\begin{center}
\includegraphics[scale=0.90]{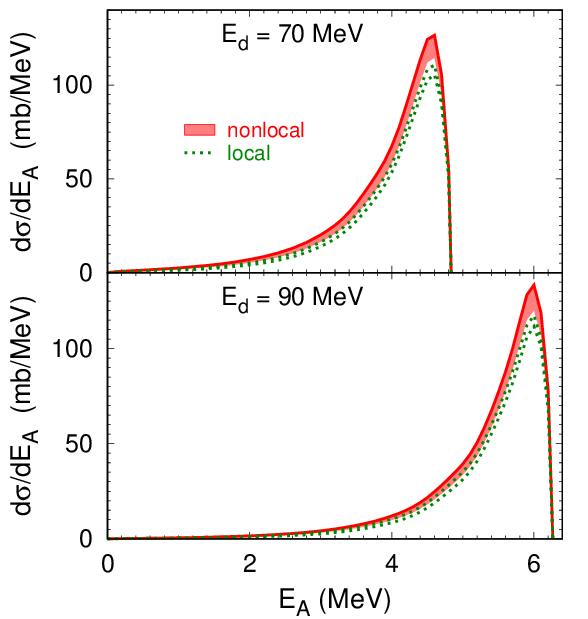}
\end{center}
\caption{\label{fig:ee} (Color online) Semi-inclusive
differential cross section for the deuteron elastic breakup on the ${}^{24}$Mg nucleus
at  $E_d = 70$ and 90 MeV as a function of the ${}^{24}$Mg energy $E_A$
in the  c.m. frame.  
Curves and bands are as in Fig.~\ref{fig:ie}.
}
\end{figure}


In Fig.~\ref{fig:e} we show the semi-inclusive
differential cross section (\ref{eq:d2s}) for the elastic breakup.
A comparison with the deuteron elastic scattering makes less
sense since the elastic differential cross section has the forward-angle Coulomb singularity
and rapid variations due to the nuclear-Coulomb interference.
The elastic  breakup cross section shows quite smooth
decrease with increasing scattering angle that is faster as compared to
the case of the inelastic breakup in Fig.~\ref{fig:i}.
The nonlocality effect remains small, slightly increasing the differential cross section
at forward angles but decreasing at larger angles.
The energy distribution presented in Fig.~\ref{fig:ee} indicates a small increase of the
cross section due to the nonlocality in the whole energy regime. This nonlocality effect
can be linked to the forward angle nonlocality effect in Fig.~\ref{fig:e}, since
  small-angle regime yields  dominating contribution to the total breakup cross section.
By comparing absolute values  in  Figs.~\ref{fig:e} and  \ref{fig:ee} with those
in Figs.~\ref{fig:i} and  \ref{fig:ie}, one notices  that
cross section for the elastic breakup significantly
exceeds cross sections for inelastic and breakup reactions with simultaneous core excitation.
 This is explicitly illustrated by integrated cross sections
for all those channels at $E_d = 90$ MeV 
collected in Table \ref{tab:s}. Noteworthy, for both the deuteron inelastic scattering and 
elastic breakup the nonlocality effect on the integrated cross section is of comparable size,
being slightly  above 10\%, but in the angular dependence it is much more pronounced in the
$(d,d')$ reaction.
The reason is that the elastic breakup exhibits a small but quite monotonic effect over a
broad kinematic regime, while in the $(d,d')$ reaction the nonlocality effect is large in
several specific regimes but alternates in sign.

\begin{table}[!h]
  \caption{Integrated cross sections (in mb) for the deuteron breakup
    and    inelastic     scattering in $d+{}^{24}$Mg collisions
    at   $E_d = 90$ MeV. Values are obtained averaging results of several local
    or nonlocal potential models, the uncertainties are evaluated as differences
between the models in each group.
}
\label{tab:s}
\centering
\begin{ruledtabular}
\begin{tabular}{*{3}{r}}
  reaction & local & nonlocal  \\ \hline
  elastic breakup & 118(5) & 132(8) \\
  inelastic breakup & 5.4(3) & 5.2(2) \\
  $(d,d')$ \, \cite{deltuva:23b} & 48.9(5) & 43.5(8) \\
\end{tabular}
\end{ruledtabular}
\end{table}

One may notice that the uncertainty due to the potential parameters, 
reflected by the band width (difference between dotted lines) for nonlocal (local) optical potentials,
is different in elastic and inelastic breakup. This might be explained partially as on-shell effect by
looking back
into $p+{}^{24}$Mg elastic and inelastic scattering in Ref.~\cite{deltuva:23b}.
The cross sections for elastic processes here and in Ref.~\cite{deltuva:23b} are less sensitive to 
local model parameters, while those for inelastic processes are less sensitive to 
nonlocal model parameters.

\section{Conclusions}

We studied the deuteron breakup on ${}^{24}$Mg nucleus. The collective excitation of the nucleus
was included  via the rotational quadrupole deformation. Rigorous three-body Faddeev-type equations
for transition operators were solved in the momentum-space framework. 
Local and nonlocal optical potentials from the previous work \cite{deltuva:23b} were
 used as dynamic input and the nonlocality effect was   investigated.

 Semi-inclusive differential cross sections with respect to angle or energy of the final
 ${}^{24}$Mg nucleus were calculated as  examples, providing a brief insight into
 the studied reactions and leading to the following conclusions:
 
 (i) The inelastic breakup is characterized by the lowest cross section,
being lower roughly by a factor of 20 than the one for the
 elastic breakup and by a factor of 10 than the one for  the inelastic deuteron scattering.

 (ii) Breakup cross sections exhibit quite small  nonlocality  effects,
  less significant than those predicted for the $(d,d')$ and $(d,p)$ reactions
 in previous works.

 (iii) The predicted  nonlocality  effects are quite different for
 elastic and inelastic breakup cross sections,  most evidently in the energy distributions,
 where the higher-energy peak
 is enhanced for the elastic breakup but suppressed for the inelastic breakup.

 Finally, our work demonstrates the feasibility of Faddeev-type calculations for
 the deuteron breakup with simultaneous collective excitation of the target nucleus
 that can be used to model reactions with other light nuclei and various interaction models,
 as well as to assess other scattering observables.

\vspace{1mm}
\begin{acknowledgments}
This work was supported by Lietuvos Mokslo Taryba
(Research Council of Lithuania) under Contract No.~S-MIP-22-72.
Part of the computations were performed using the infrastructure of
the Lithuanian Particle Physics Consortium.
\end{acknowledgments}


 \end{document}